\documentclass[iop,numberedappendix]{emulateapj-rtx4}
\usepackage{mathrsfs }
\usepackage{natbib}
\usepackage{multirow}
\usepackage{amssymb}
\usepackage{ulem} 
\newcommand{\msun}{\ensuremath{M_\odot}}

\newcommand{\NHI}{\ensuremath{N_{\rm HI}}}

\def\kms{km~s$^{-1}$}
\def\cm2{cm$^{-2}$}
\def\Msun{$M_{\odot}$}

\def\ergscm2{$\rm erg~ \rm s^{-1} \rm ~cm^{-2} $}

\usepackage{float}

\shorttitle{A Unique View of AGN-Driven Molecular Outflows}
\shortauthors{Rudie et al. }

\begin{document}

\title{A Unique View of AGN-Driven Molecular Outflows: The Discovery of a Massive Galaxy Counterpart to a $z=2.4$ High-Metallicity Damped Lyman-$\alpha$ Absorber}

\author{Gwen C. Rudie, Andrew B. Newman,}
\affiliation{The Observatories of the Carnegie Institution for Science \\
813 Santa Barbara Street \\
Pasadena, California 91101, USA}

\and

\author{Michael T. Murphy}
\affiliation{Centre for Astrophysics and Supercomputing\\
Swinburne University of Technology\\
Hawthorn, VIC 3122, Australia}

\begin{abstract}
We report the discovery of a massive $\log(M/M_\odot)=10.74^{+0.18}_{-0.16}$ galaxy at the same redshift as a carbon-monoxide-bearing sub-damped Lyman $\alpha$ absorber (sub-DLA) seen in the spectrum of QSO J1439+1117. The galaxy, J1439B, is located 4\farcs7 from the QSO sightline, a projected distance of 38 physical kpc at $z=2.4189$, and exhibits broad optical emission lines ($\sigma_{\rm{[O III]}}=303 \pm 12$~\kms) with ratios characteristic of excitation by an active galactic nucleus (AGN). The galaxy has a factor of $\sim$9 lower star formation than is typical of star-forming galaxies of the same mass and redshift. The nearby sub-DLA is highly enriched, suggesting its galactic counterpart must be massive if it follows the $z\sim2$ mass-metallicity relationship. Metallic absorption within the circumgalactic medium of the sub-DLA and J1439B is spread over a velocity range $\Delta v > 1000$ \kms, suggesting an energetic origin. We explore the possibility that a different galaxy could be responsible for the rare absorber, and conclude that it is unlikely based on imaging, integral-field spectroscopy, and high-$z$ massive galaxy pair statistics. We argue that the gas seen in absorption against the QSO was likely ejected from the galaxy J1439B and therefore provides a unique observational probe of AGN feedback in the distant universe. 
\end{abstract}


\section{Introduction}

The formation and evolution of galaxies is inextricably tied to the flow of gas into and out of galaxies through the circumgalactic medium (CGM). Galaxies in the early universe require significant gas accretion in order to power their high rates of star formation for sizable fractions of a Hubble time (see, e.g., \citealt{erb08} and  \citealt{fin08}). Gas outflows are observationally known to be prevalent among high-redshift galaxies \citep{pet01, sha03, ccs10}, and cosmological semi-analytic and hydrodynamic simulations require strong feedback in order to reproduce the observed luminosity function of galaxies \citep{whi91,col94,kau99,som99,efs00}. At low masses, the energy and momentum injected into the gas from supernovae and stellar radiation pressure is likely sufficient to drive galaxy-scale winds \citep{mur11,hop12}; however, at masses above $M^*$, feedback from active galactic nuclei (AGN) may be needed to regulate and eventually halt star formation \citep{ben03,cro06}.

The gaseous environments of high-redshift galaxies are an excellent laboratory to explore the properties of galactic accretion and winds. Studies of statistical samples of high-redshift galaxies have begun to quantify the properties of the CGM \citep{gcr12,tur14,tur15,rub15,lau16}; however, there is still much observational and theoretical progress needed to understand the nuanced relationship between galaxies and their gaseous surroundings, especially with regard to feedback. 

\begin{figure*}
\center
\includegraphics[width=\textwidth]{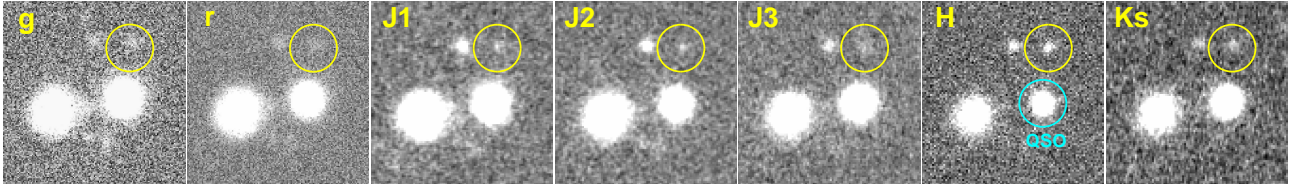}
\caption{Optical and NIR photometry of the galaxy 4\farcs7 from the background $z=2.58$ QSO. Each panel is 15" on a side. North is up and east is to the left. QSO J1439+1117 is circled in cyan in the $H$-band panel. The galaxy J1439B is circled in yellow in all panels. The photometry of the other objects detected in these images suggest they are at a lower redshift than J1439B ($z_{\rm{phot}}<1.9$). }
\label{images}
\end{figure*}

One particular class of CGM absorption is known as the damped Lyman $\alpha$ absorber (DLA). Covering about a third of the $2<z<3$ sky \citep{fum15},  DLAs contain the majority of the neutral gas at all redshifts \citep{tyt87,ome07} and are thought to trace gas within the interstellar medium (ISM) or very near to galaxies. Compared to the rest of the intergalactic medium (IGM) and CGM which is predominantly ionized, DLAs have column densities of hydrogen large enough that the outer layers absorb the incident hydrogen-ionizing photons, leaving the majority of the gas within the structure neutral. For historical reasons, DLAs are usually defined as those clouds with $\log(N_{\rm{HI}})>20.3$ (see \citealt{wol05} for a review), although structures with somewhat lower column densities are still self-shielding and may be largely neutral \citep{des03,mei09,leh14}.  

DLAs, like galaxies, have a wide variety of physical properties that likely reflect the diversity of their origins. The majority of DLAs are metal poor \citep{pet97,raf12,jor13}, have small physical sizes \citep{mon09,coo10}, and are kinematically cold \citep{pro97,nee13}. Although some authors have suggested that they could originate in the far outskirts of more massive systems \citep{coo06,raf11,fon12}, both cosmological and high-resolution simulations suggest that the majority of DLAs are associated with relatively low-mass galaxies  \citep{hae98,pon08,bir15}.  

Tremendous effort has been focused on the identification of the galaxy counterparts of high-redshift DLAs. Large surveys of the volume surrounding QSO sight lines with DLAs have been carried out using both long-slit \citep{fyn10} and integral-field \citep{per11,bou12,per12} spectrographs as well as imaging and spectroscopic surveys using double DLA sight lines \citep{fum10}. Despite these efforts, however, relatively few galaxy counterparts are known. \citet{per11}, \citet{bou12}, and \citet{fum15} employed systematic searches for DLA host galaxies. With relatively few detections, based on the detection limits of their data, the authors conclude that the typical DLA galaxy counterpart has a star formation rate SFR~$<1-3$ \Msun~yr$^{-1}$. In contrast, the detected galactic counterparts to $z>2$ DLAs unsurprisingly represent a biased sample; the galaxies typically have significantly higher SFRs, and on average, the metallicity of the DLAs is higher \citep{fyn10, per12, fum15}

\citet{fum15} recently reviewed the published results and found that only 11 $z>1.9$, $\log(N_{\rm{HI}})>20.3$ absorption systems were known to have galactic counterparts at the same spectroscopically identified redshift \citep{mol93,mol02,mol04,wea05,fyn10,fyn11,bou12,kro12,not12,per13,bou13,jor14}. Since 2015, at least five additional plausible host galaxies have been identified \citep{har15,sri16,nee17}. In addition to these, a handful of high-$z$ sub-DLAs with $19.0<\log(N_{\rm{HI}})<20.3$ have identified nearby galaxies (c.f. \citealt{kas14,zaf17}). Compared with either the statistical sample of DLAs from the Sloan Digital Sky Survey, which contains over 12,000 $z>2$ DLAs \citep{not12b}, or the several hundred DLAs with high-resolutions spectroscopy (see, e.g., \citealt{zaf13,jor13}), the sample of galaxy counterparts is remarkably small. Given the current small sample, additional discoveries are critical both to our understanding of the nature of DLAs and sub-DLAs, as well as to our understanding of galaxy formation. 

The sub-DLA at z=2.41827 toward QSO J1439+1117, hereafter DLA$_{\rm{J1439}}$, is a unique system due to its high level of chemical enrichment and molecular gas content. It has an \ion{H}{1} column density $\log(N_{\rm{HI}})=20.10\pm0.10$, placing it just short of the classical DLA limit of $\log(N_{\rm{HI}}) \ge 20.3$, and metallicity consistent with the solar value \citep{not08}.  Gas associated with the sub-DLA is detected in numerous metallic transitions including neutral carbon and sulfur, as well as in several molecular species (see \citealt{not08}, \citealt{sri08}, and Section \ref{QSO}), suggesting that the absorber is self-shielding. DLA$_{\rm{J1439}}$ provided the first optical detection of carbon monoxide (CO) associated with a DLA or sub-DLA \citep{sri08}. In addition to CO, H$_{\rm {2}}$ and HD are also detected \citep{sri08,not08}. Further, DLA$_{\rm{J1439}}$ has a high molecular fraction $f=2N(\rm{H}_2)/(N(\rm{H}~ I)+2N(\rm{H}_2))=0.27^{+0.10}_{-0.08}$ \citep{sri08}, the highest measured in any DLA or sub-DLA to date \citep{lis15}.

Here we report the discovery of a galaxy at the same redshift $z=2.4189$ separated from DLA$_{\rm{J1439}}$ by 4\farcs7, or 38 physical kpc. In Section \ref{data}, we describe the imaging and spectroscopic discovery data and their analysis. Section \ref{J1439B} describes the properties of J1439B that indicate it likely hosts an AGN and which make it a likely galaxy counterpart to DLA$_{\rm{J1439}}$. In Section \ref{discussion}, we argue that J1439B is the most likely candidate for the source of the absorber and discuss the implications for the chemical evolution of the galaxy and for AGN-driven outflows. 

Throughout this paper we assume a \citet{cha03} stellar initial mass function and a $\Lambda$-CDM cosmology with $H_{0} = 70$ \kms\  Mpc$^{-1}$, $\Omega_{\rm m} = 0.3$, and $\Omega_{\Lambda} = 0.7$.  Unless otherwise specified, all distances are in physical units, all transitions are referred to by their vacuum wavelengths, and magnitudes refer to the AB system \citep{ABmag}.

\section{Data}
\label{data}

\subsection{QSO Spectroscopy}

High-resolution spectroscopic observations of QSO J1439+1117 were carried out using the Ultraviolet and Visual Echelle Spectrograph (UVES; \citealt{UVES}) on the Very Large Telescope (VLT). The data from ESO program 278.A-5062(A) were obtained from the ESO Science Archive Facility. A description of the observational setup is discussed in \citet{not08} and \citet{sri08}. All column density measurements of the absorption systems, and quantities derived from them,  are taken from these papers, and the QSO data are presented in this work for visualization of the velocity distribution only.

A detailed description of the reduction procedure can be found in \citet{mur07} and \citet{bag14}. Briefly, the data were reduced using the ESO UVES pipeline. Cosmic rays were masked, and the exposures were combined using UVES\_popler \citep{mur16} which was specifically designed to optimally combine exposures reduced by the UVES pipeline. The resultant spectrum was cleaned of artifacts and continuum normalized.

\subsection{Imaging Observations}

Near-infrared (IR) imaging of the field surrounding QSO J1439+1117 was taken with the FourStar camera \citep{fourstar} on the Magellan Baade 6.5~m telescope.  The FourStar data were reduced using a custom pipeline, FourCLift, developed by D. Kelson and described in \citet{fourclift}. FourCLift corrects for dark current and nonlinearity before flat-fielding the images. Bad pixels are masked, and the sky background is determined iteratively using a bivariate wavelet transformation of the images with detected sources masked. Time variability in each level of the wavelet transforms is determined and sky frames are computed by inverting model wavelet transforms reconstructed for the time of each science exposure. These model sky frames are subtracted, and the frames subsequently aligned and stacked, taking into account the distortion of the camera to produce rectified images. Photometric calibration of the FourCLift processed images was performed with unsaturated 2MASS stars in the field. 

\begin{figure}
\center
\includegraphics[width=0.5\textwidth]{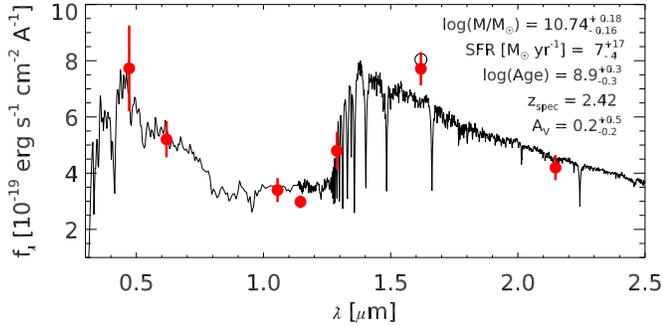}
\caption{ Aperture and foreground dust-corrected photometry of the galaxy J1439B is shown in red. The red $H$-band data point shows the photometry corrected for the [\ion{O}{3}] $\lambda\lambda5008,4960$ line flux, while the black open circle shows the photometry without this correction. In black is plotted the best-fit SED from FAST. The parameters of the best-fit model are listed in the upper right-hand corner.}
\label{sed}
\end{figure}

Optical images in Sloan $g'$ and $r'$ were taken with the $f/2$ camera of the IMACS imaging spectrometer \citep{imacs} on the Magellan Baade telescope. The data were bias and gain corrected, flat fielded, background subtracted, and combined using standard IRAF packages. The zeropoints of the images were determined using photometrically calibrated stars in the same fields from the SDSS DR7 \citep{sdss,SDSS_DR7}. The optical and NIR images are shown in Figure \ref{images}. All of the images were aligned and then registered to the $H$-band image. They were then smoothed to a FWHM = 1\farcs2, similar to that of the $K_s$ band image.

\begin{figure*}
\center
\includegraphics[width=\textwidth]{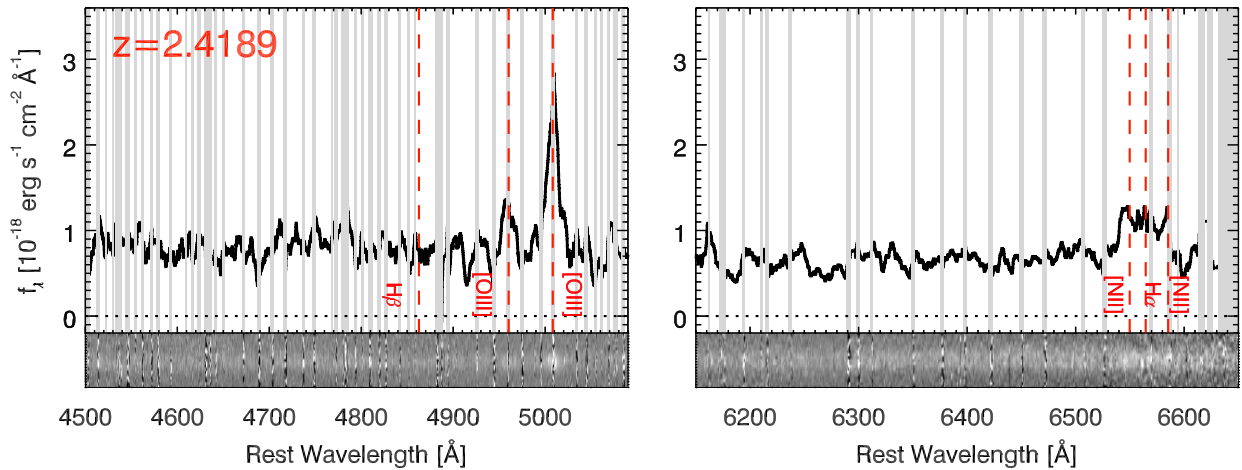}
\caption{ $H$ (left) and $K$ (right) band portions of the FIRE spectrum of the galaxy. The top panels show the smoothed 1D spectrum with relevant emission lines marked. The bottom panel shows the 2D spectrum. Gray bands indicate portions of the spectrum that are contaminated by night-sky emission and are masked during analysis.}
\label{HKspec}
\end{figure*}

\subsection{Galaxy Photometry}

From the NIR imaging, a galaxy of interest (henceforth J1439B) was identified 4\farcs7 north of the QSO at R.A. $14^h39^m12^s.0$, decl. $+11^{\circ}17'45''$. The colors of the galaxy were measured using Source Extractor \citep{sourceextractor} in a 1\farcs6 aperture centered on the $H$-band detection in order to minimize any contribution from the nearby QSO. A small correction to the colors for light that falls outside of the 1\farcs6 aperture was computed and applied by comparing the color of stars measured in 8" apertures to that measured in 1\farcs6 apertures.  Conservatively, the color errors were computed as the quadratic sum of the photometric error in the two bands and the aperture correction. 

The $H$-band photometry of the galaxy was computed using the un-smoothed image. The Source Extractor $H$-band MagAuto aperture was used as the normalization and the magnitudes in the other photometric bands were computed by adding the $H$-band magnitude to the aperture-corrected colors. The magnitude errors used throughout this paper are the quadratic sum of the aperture-corrected color errors and the $H$-band MagAuto error computed by Source Extractor. The final magnitudes, corrected for Galactic extinction following \citet{sch11}, are reported in Table \ref{phot} and shown in Figure \ref{sed}. 

\begin{deluxetable}{lcc}
\tablecaption{ Galaxy Photometry}  
\tablewidth{0pt}
\tablehead{
\colhead{Band} & \colhead{Instrument} &  \colhead{AB Magnitude\tablenotemark{a} } \\
\colhead{} & \colhead{} &  \colhead{} }
\startdata
$g$ & IMACS & 24.50  $\pm$ 0.21 \\
$r$ & IMACS &  24.34  $\pm$  0.13 \\
$J1$ & FourStar & 23.65  $\pm$ 0.14 \\
$J2$ & FourStar & 23.61 $\pm$ 0.08  \\
$J3$ & FourStar & 22.84  $\pm$   0.15  \\
$H$ & FourStar & 21.78 $\pm$  0.07 \\
$H_{\rm corr}$\tablenotemark{b} & FourStar & 21.83  $\pm$  0.08\\
$K_s$ & FourStar &  21.87  $\pm$     0.12\\
 \enddata
     \label{phot}
     \tablenotetext{a}{Aperture and foreground dust-corrected magnitudes and errors.}
     \tablenotetext{b}{The $H$-band magnitude corrected for the [\ion{O}{3}]$\lambda$5008 and $\lambda$4960 emission as determined from the FIRE observations.}
\end{deluxetable}

\subsection{NIR Spectroscopy}

An NIR spectrum of the galaxy was acquired in 0\farcs6 seeing with the FIRE spectrograph \citep{fire} on the Magellan Baade telescope. In total, 7 hours of integration were obtained using 20-minute exposures taken in an AB dither pattern. Preliminary reduction of the data was completed with the publicly available reduction pipeline FIREHOSE.\footnote{ Written by Rob Simcoe, John Bochanski, and Mike Matejek; http://www.firespectrograph.org/} The flat field was derived from internal quartz flats and the illumination correction was computed from twilight flat exposures. The initial wavelength solution from FIREHOSE utilizes both ThAr arc exposures as well as OH emission lines captured in the science exposures. The data were also flux calibrated and corrected for telluric absorption using observations of A0V stars. 

Custom reduction software was developed to produce a rectified 2D spectrum (see \citealt{new15}). The trace of the object through the echelle orders was mapped using observations of telluric calibration stars, and each echellogram was rectified. Spatial object profiles were derived from the spectral regions surrounding the [\ion{O}{3}] emission lines detected in the FIREHOSE-derived sky-subtracted combined exposures for the A and B positions from each night. These profiles were used to model the contributions to the 2D spectrum from the source for second-pass sky subtraction. The wavelength solution was refined outside of FIREHOSE using OH lines in the observed data. Individual exposures were aligned and averaged without weighting, and the 1D spectrum was extracted with a boxcar of 1\farcs0 width. The resultant spectrum was renormalized to match the $H$-band photometry and the best-fit spectral energy distribution (SED) template in order to correct for possible slit losses and errors in the spectrophotometry. The reduced spectrum is shown in Figure \ref{HKspec}.

\begin{figure*}
\center
\includegraphics[width=\textwidth]{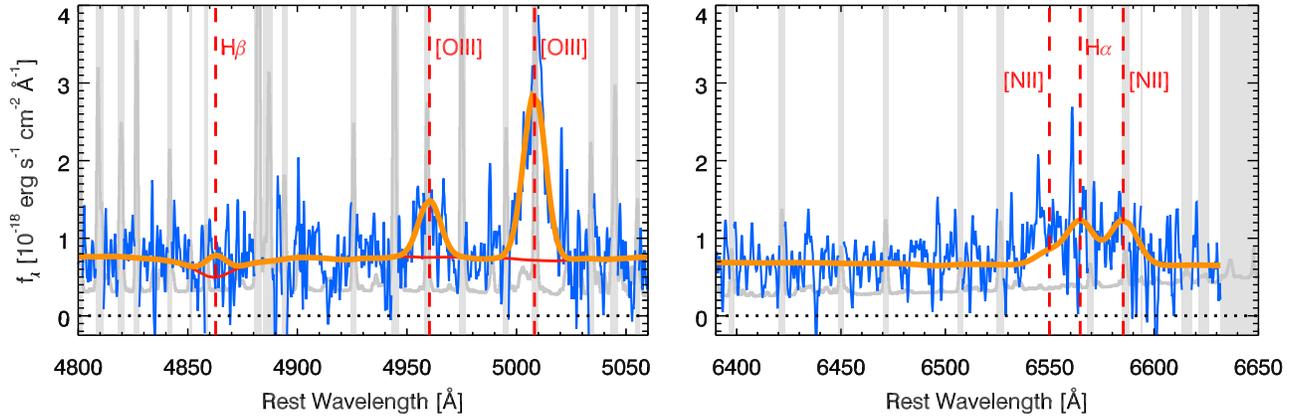}
\caption{Gaussian fits to the 1D spectrum in $H$ (left) and $K$ (right) bands. The data are shown in blue, along with their 1$\sigma$ errors in dark gray. Regions near night-sky lines are masked in light gray and excluded from the fit. The fit is shown in orange. In the $H$ band, the normalized best-fit SED model used as the stellar continuum in the fit (see the text) is shown in red.  }
\label{fit}
\end{figure*}

\subsection{Photometric Modeling}

The SED fitting code FAST \citep{fast} was used to model the photometry of the galaxy using the \citet{bc03} stellar populations synthesis models, a \citet{cha03} IMF, and a \citet{cal00} dust-attenuation curve. As the spectral library does not include emission lines, we also correct the $H$-band photometry by 0.045 mag to account for the presence of  [\ion{O}{3}] $\lambda\lambda5008,4960$ line emission, the strongest line present in the FIRE spectrum (see Section \ref{spec_fit}). The redshift was fixed to the measured spectroscopic redshift of the galaxy $z_{\rm {spec}}=2.42$. The best-fit model derived by FAST implies that the galaxy is relatively massive, $\log(M/M_\odot)=10.74^{+0.18}_{-0.16}$, with a comparatively modest star formation rate, $7^{+17}_{-4}$~\Msun~yr$^{-1}$, and a relatively evolved stellar population (age~$=700^{+700}_{-300}$~Myr). The photometric data as well as the best-fit SED are shown in Figure \ref{sed}. 

\label{SED_text}

\vspace{1cm}

\subsection{Spectroscopic Modeling}

\label{spec_fit}

Rest-optical spectroscopy of the galaxy (Figure \ref{HKspec}) reveals broad [\ion{O}{3}] emission in the $H$ band and blended H$\alpha$ and [\ion{N}{2}] emission in the $K$ band. The $H$-band portion of the spectrum was modeled using a single Gaussian emission component for each of the [\ion{O}{3}] $\lambda\lambda5008,4960$ and H$\beta$ emission lines superimposed on the best-fit stellar continuum model from the SED fitting. The best-fit SED model spectrum was convolved with a Gaussian matched to the measured [\ion{O}{3}] emission line width ($\sigma=300$ \kms). This model continuum was used to account for stellar absorption at H$\beta$. The emission lines were all forced to have the same redshift and velocity dispersion, and the [\ion{O}{3}] emission line fluxes were forced to have a  [\ion{O}{3}] $\lambda5008/\lambda4960$ ratio of 2.98, in agreement with their theoretically determined  magnetic-dipole transition probabilities \citep{sto00}. Regions around bright night-sky lines were masked prior to fitting (gray regions in Figure \ref{fit}). The resulting fit is shown in the left-hand panel of Figure \ref{fit} and the parameters of the fit are listed in Table \ref{specfit}.

Both [\ion{O}{3}] lines are well-detected and best fit by a broad Gaussian with $\sigma=303\pm12$ \kms\ at $z=2.4189\pm0.0001$. Using the best-fit SED spectrum as the model continuum yields a formal $4.8\sigma$ detection of H$\beta$ and a high value of [\ion{O}{3}]/H$\beta = 8.0 \pm 1.7$.\footnote{If the stellar continuum were modeled without the expected stellar absorption, H$\beta$ would be undetected leading to a higher value of [\ion{O}{3}]/H$\beta$.} However, given the uncertainty in the precise age of the stars and the stellar velocity dispersion of the galaxy, and therefore the shape and normalization of the underlying H$\beta$ absorption, we consider one additional model to determine the maximum allowable H$\beta$ line flux and the lowest possible value of [\ion{O}{3}]/H$\beta$. We fix all of the input parameters in FAST to the best-fit values, but set the age to $10^{8.5}$ year, and a short exponential star-formation timescale $\tau=10^7$ year in order to produce a spectrum with the largest plausible stellar absorption at H$\beta$.\footnote{The \citet{bc03} models have a maximum H$\beta$ absorption equivalent width at an age of $10^{8.5}$ year.} Fitting the $H$-band spectrum using this maximal-H$\beta$ model as the continuum results in a fit with higher H$\beta$ line flux; however, the change is smaller than the error in the fit using the best-fit SED. 

[\ion{N}{2}]$\lambda\lambda6549,6585$ and H$\alpha$ emission is also detected at lower significance and is more difficult to model robustly. Because of this we use a model similar to the one above, with single Gaussian components for [\ion{N}{2}] and H$\alpha$ superimposed on the stellar continuum; however, we force the redshift and velocity dispersion of the emission lines to match that determined by the $H$-band fit (Figure \ref{fit}, right panel). As with [\ion{O}{3}], we fix the ratio of the [\ion{N}{2}] emission lines to their theoretically determined [\ion{N}{2}] $\lambda6585/\lambda6549$ ratio of 3.05. This results in a fit that appears to under-predict the emission at wavelengths shorter than that of H$\alpha$. The detected excess emission could be due to additional H$\alpha$ emission, which is blue-shifted with respect to the main H$\alpha$ component; however the quality of the $K$-band data do not warrant a multi-component fit to this emission. The parameters of both the $H$- and $K$-band fits are listed in Table \ref{specfit}.

\begin{deluxetable*}{lc|cccc|cc}
\tablecaption{Emission Line Fit Parameters}  
\tablewidth{0pt}
\tablehead{
\colhead{Redshift} & \colhead{$\sigma_{\rm[OIII]}$ } &   \multicolumn{4}{c}{Line Flux ($10^{-18}$ erg s$^{-1}$ cm$^{-2}$)}  & \multicolumn{2}{c}{Line Ratios}\\
\colhead{} & \colhead{(km s$^{-1}$) } &  \colhead{[\ion{O}{3}]$\lambda5008$} &  \colhead{H$\beta$\tablenotemark{a}} &  \colhead{H$\alpha$} &  \colhead{[\ion{N}{2}]$\lambda6585$} & \colhead{[\ion{O}{3}]$\lambda5008$/H$\beta$\tablenotemark{a}}  & \colhead{[\ion{N}{2}]$\lambda6585$/H$\alpha$} }
\startdata
 $2.4189 \pm   0.0001$& $303 \pm 12$& $93.5 \pm   3.5$ & $ 11.7 \pm   2.4$  & $38.6 \pm   3.7$ & $34.2 \pm   4.0$  & $8.0 \pm 1.7$ & $0.88 \pm      0.13$
 \enddata
     \label{specfit}
     \tablenotetext{a}{This fit reflects the use of the best-fit SED as the continuum in order to correct for underlying stellar absorption. Without this correction, H$\beta$ is undetected. }
    \end{deluxetable*}

\section{The Nature of J1439B}

\label{J1439B}

Given the close separation (4\farcs7, 38 physical kpc) of the galaxy J1439B to the high-metallicity, CO-bearing sub-DLA seen in the spectrum of QSO J1439+1117, the properties of the galaxy are of considerable interest. In particular, it is necessary to assess the possibility that the galaxy could have a causal relationship with the sub-DLA in which the gas represents the ejected or stripped ISM of J1439B. Given this, signatures of past feedback within the galaxy, as well as its present capability to launch gas, are of relevance. As will be shown, the galaxy sits below the main sequence of star formation, has broad nebular line emission plausibly due to an outflow, appears likely to host an AGN, and is massive enough to have enriched its ISM to near solar metallicity, consistent with the abundances derived for DLA$_{\rm{J1439}}$. Collectively, these properties suggest that J1439B is the likely source of the sub-DLA gas. 

\subsection{The Position of J1439B with Respect to the Main Sequence of Star Formation}

\label{MSSF_text}

\begin{figure}
\center
\includegraphics[width=0.5\textwidth]{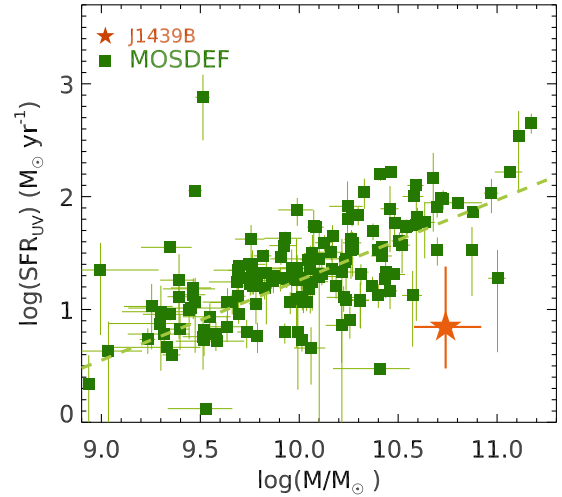}
\caption{The SFR-stellar mass relation for $z\simeq2$ galaxies. In green squares are shown galaxies with $2.0<z<2.6$ from the MOSDEF sample from \citet{shi15}. The plotted values are SFRs determined from the dust-corrected UV luminosity at 1600\AA\ that are more analogous to the SED-determined SFR for J1439B.  The same qualitative picture would hold if H$\alpha$-based SFRs were used for the MOSDEF sample. The green dashed line shows the fit to the main sequence from \citet{shi15} as described in the text. The galaxy J1439B is plotted as the red star. Compared with typical star-forming galaxies from MOSDEF, at fixed stellar mass, J1439B exhibits significantly lower levels of star formation. Using the main sequence fit and measured scatter ($\sigma_{\rm{MSSF}}$ ) from \citet{shi15}, J1439B lies $3.8^{+1.5}_{-2.1} \sigma_{\rm{MSSF}}$ below the main sequence. }
\label{MSSF}
\end{figure}

\begin{figure*}
\center
\includegraphics[width=0.5\textwidth]{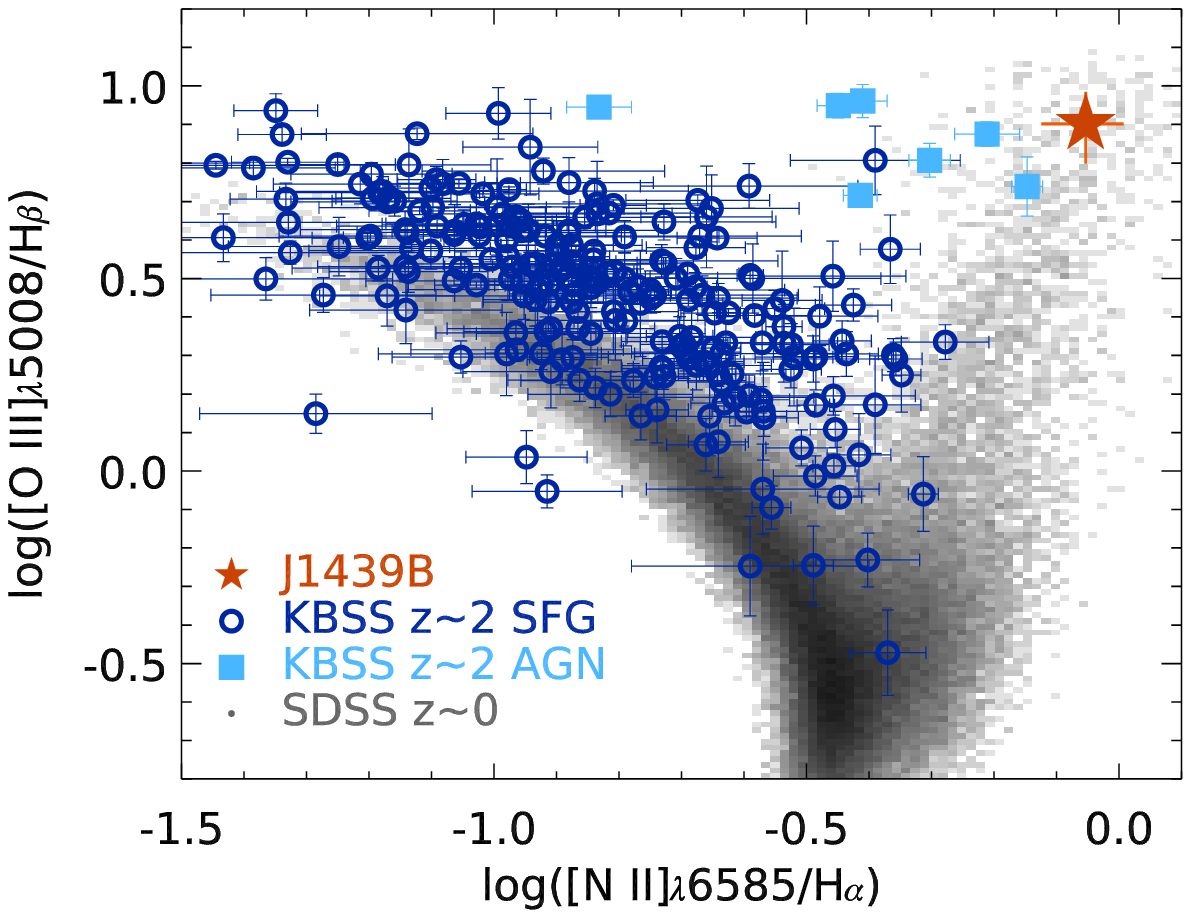}\includegraphics[width=0.5\textwidth]{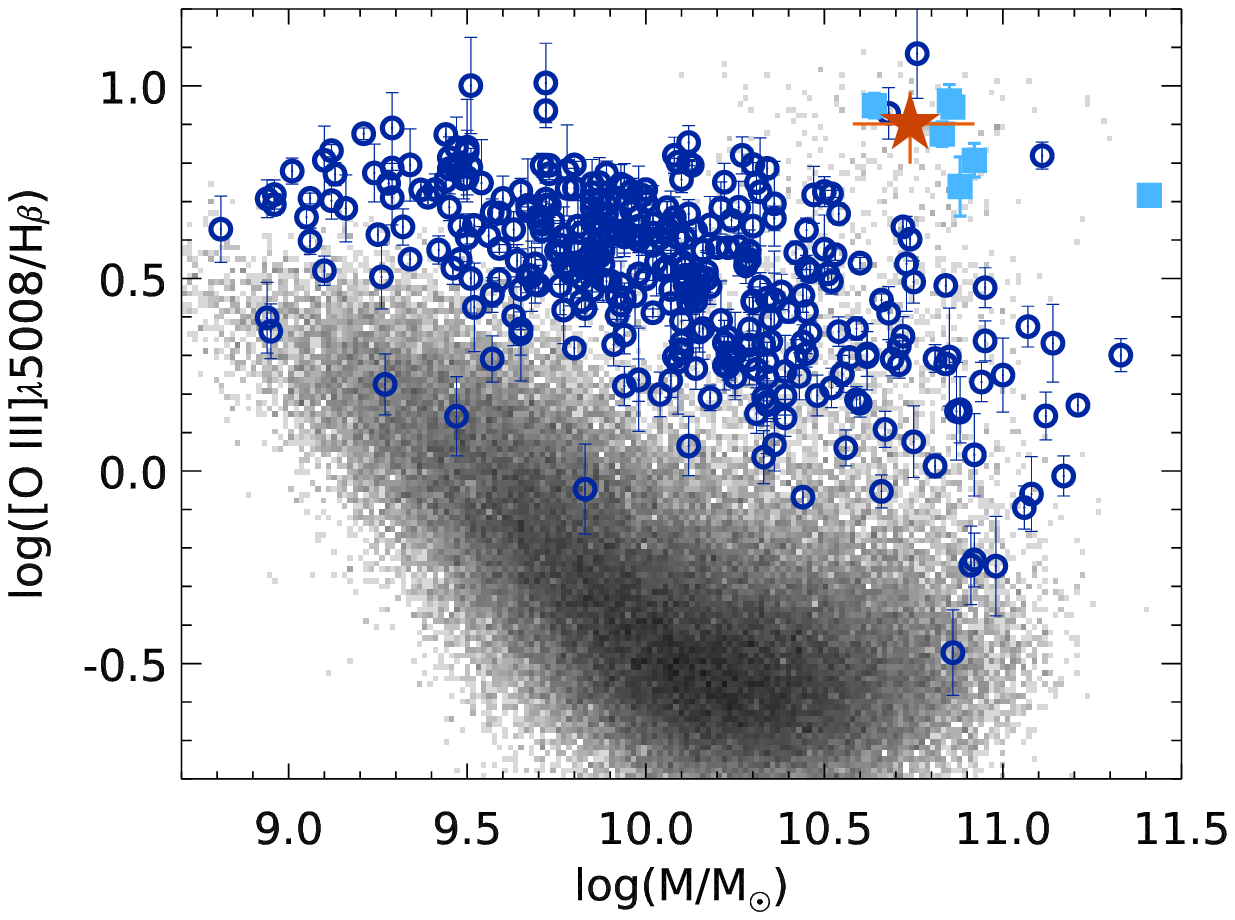}
\caption{ The BPT (left) and mass-excitation (MEx, right) diagrams for $z\simeq0$ and $z\simeq2$ galaxies. Low-redshift galaxies from the SDSS are shown in gray. In blue circles are plotted $z\simeq2$ star-forming galaxies from the KBSS \citep{str17}. Galaxies spectroscopically identified to be AGN within the KBSS are plotted as blue squares. The galaxy J1439B is represented by the red star. J1439B has very high excitation (high [\ion{O}{3}]/H$\beta$), among the highest in the KBSS sample. Given its high mass and high value of [\ion{O}{3}]/H$\beta$ and likely [\ion{N}{2}]/H$\alpha$, the galaxy J1439B likely harbors an AGN. }
\label{BPT}
\end{figure*}

Many authors have noted that star-forming galaxies exhibit a correlation between their stellar mass and star-formation rate such that high-mass galaxies typically exhibit higher levels of star formation, both in the local universe and at high redshift \citep{bri04,noe07,dad07,rod11,wuy11,red12,whi12}. Generally referred to as the ``main sequence'' of star formation following \citet{noe07}, the underlying meaning of this relation and its evolution is debated (see e.g. \citealt{pen10,kel14, abr14}); however, the existence of such a correlation allows one to infer how ``typical" a given SFR is for a galaxy at fixed stellar mass and redshift. 

Shown in Figure \ref{MSSF} is a recent measurement of the main sequence of $z>2$ galaxies from the $H$-band selected MOSDEF sample as presented in \citet{shi15}.\footnote{ \citet{shi15} measure a completeness limit of $M_* = 10^{9.5}$ M$_\odot$. For galaxies with masses comparable to J1439B, the sample should be highly complete.} As we determine the SFR of J1439B based on the best-fit SED, the values plotted from the MOSDEF sample are those based on photometrically derived SFRs; however, the results are unchanged if one considers H$\alpha$-based SFRs. Also plotted for comparison is the SED-based SFR and stellar mass of J1439B.\footnote{We argue below that the dominant ionization source in J1439B is an AGN, and so we refrain from reporting an SFR determined from H$\alpha$ emission. The SED-derived SFR implies significantly less flux in the H$\alpha$ emission line than is measured, consistent with the argument that the SED is mostly determined by the stellar properties while the ionizing emission lines are dominated by an AGN.} 

It is evident from this figure that J1439B has a low rate of star formation given its present stellar mass compared to galaxies from the MOSDEF sample. To better quantify how ``atypical'' J1439B is, we use the fit to the main sequence described by \citet{shi15} to galaxies from the MOSDEF sample with $2.09<z<2.61$ using the UV SFR indicator. They found
\begin{equation}
\log(\rm{SFR}) = [0.71 \pm 0.09] \times \log(M_*) -5.84\pm0.92
\end{equation} 
with an intrinsic scatter of $\sigma_{\rm{MSSF}}=0.25$ where SFR is measured in \Msun\ yr$^{-1}$ and $M_*$ is measured in \Msun. Based on this fit, for a galaxy with the mass of J1439B, we would expect an SFR of 61 \Msun\ yr$^{-1}$. The SFR of J1439B, SRF$=7^{+17}_{-4}$, is therefore $3.8^{+1.5}_{-2.1} \sigma_{\rm{MSSF}}$ below the main sequence determined by \citet{shi15}.

\subsection{The Ionization Properties of J1439 from Optical Emission Line Ratio Diagnostics}

\label{BPT_text}

The deep rest-optical spectrum of J1439B provides important constraints on the source of ionization in the galaxy. One of the best known diagnostics, the [\ion{N}{2}] BPT diagram \citep{bal81}, is shown in the left-hand panel of Figure \ref{BPT}. Local galaxies, taken from the SDSS Data Release 7 \citep{SDSS_DR7} are shown in the gray shading. The high-density arc of galaxies extending from the bottom middle of the plot toward the upper left is a sequence of low-$z$ star-forming galaxies in which [\ion{O}{3}]/H$\beta$ increases and  [\ion{N}{2}]/H$\alpha$ decreases with decreasing gas-phase metallicity and increasing ionization. Galaxies that lie in the more diffuse distribution have optical line emission that is dominated by ionization from an AGN. The region in between is populated by galaxies with both AGN and star-formation, or those with ionization that is dominated by shocks or hot evolved stars \citep{vei87,kew01,kau03}. 

Overplotted in blue are a large sample of UV-selected $z\simeq2.3$ star-forming galaxies from the Keck Baryonic Structure Survey (KBSS; \citealt{str17}). From the KBSS sample, we plot galaxies with $>5\sigma$ measurements of H$\alpha$, $>3\sigma$ measurements of [\ion{O}{3}]$\lambda5008$ and H$\beta$, and $>2\sigma$ measurements of [\ion{N}{2}]$\lambda6584$. Galaxies that are spectroscopically identified as AGN are plotted in light blue squares, while galaxies without detected spectroscopic signatures of AGN activity are plotted as dark blue open circles. As described in \citet{ccs14} and \citet{str17},  galaxies are identified as AGN based on strong detections of high-ionization emission lines in their rest-UV spectrum (e.g. \ion{N}{5}$\lambda1240$, \ion{C}{4}$\lambda1549$)  or, for those galaxies lacking rest-UV spectroscopy,\footnote{Two out of the seven galaxies identified as AGN in \citet{ccs14} lack UV spectroscopy.} unambiguous combinations of rest-optical line widths and ratios characteristic of AGN. 

The location of J1439B within this diagram is represented by the red star. As had been discussed extensively in the literature \citep{mas14,ccs14, sha15,san16,str17}, based on large high-redshift galaxy samples, there is clearly a systematic offset between the $z\simeq2$  and $z\simeq0$ star-forming loci. Given the higher level of ionization of high-$z$ galaxies, and the possibility of metal-poor AGN, it is likely that the [\ion{N}{2}] BPT diagram no longer provides a clean separation for all galaxies and AGN \citep{kew13}. Based on the MOSDEF sample, which includes IR- and X-ray-selected AGN, \citet{coi14} and \citet{aza17} argue that some AGN may contaminate galaxies in the star-forming sequence of the BPT diagram, but that galaxies at the plotted position of J1439B are almost certainly AGN. Considering Figure \ref{BPT}, based on the location of either low- or high-redshift AGN, the position of J1439B is quite clearly within the AGN portion of the diagram.  However, caution is still warranted given the lower S/N of the FIRE spectrum in the $K$ band and the resulting poor emission line model fit. The measured ratio of [\ion{N}{2}]$\lambda6585$/H$\alpha$ is higher than all of the KBSS $z\simeq2$ galaxies; however, the measurement is likely subject to additional systematic uncertainties.\footnote{The plotted error is based solely on the errors in the fit parameterized as discussed in Section \ref{spec_fit}. No source of systematic error due to the choice of model is taken into account.}

To diagnose the likely presence of an AGN without relying on the comparatively poor quality of the $K$-band spectrum, we also consider a mass-excitation diagram (MEx; \citealt{jun11}) in the right-hand panel of Figure \ref{BPT}. The same samples are plotted as described above, but the requirements of a detection of [\ion{N}{2}] among the KBSS sample is relaxed. As was the case with the BPT diagram, the position of J1439B in the MEx diagram is most similar to the most extreme AGN in the low-$z$ universe and coincident with all of the high-$z$ AGN from the KBSS. While neither of these diagnostics is completely conclusive as to the nature of the ionization source, it appears likely that the galaxy hosts an AGN.

\subsection{Emission Line Width}

\label{width_text}

Another distinguishing property of the optical spectrum of J1439B is the large velocity dispersion $\sigma=303 \pm 12$~\kms~measured in the [\ion{O}{3}] emission lines. This value of $\sigma$ is larger than all of the galaxies in the UV-selected KBSS-MOSFIRE sample\footnote{The only galaxy with a comparable velocity dispersion is Q0821-BX101, which \citet{ccs14} identified as an AGN.} \citep{str17}, even those galaxies with significantly higher inferred stellar mass (see Figure \ref{m-sigma}). Measured values of $\sigma \gtrsim 300$~\kms~ do exist among high-redshift samples, most of which fall into two distinct types: compact star-forming (c.f. \citealt{van15}) and quiescent (c.f. \citealt{van09,bel17}) galaxies, as well as AGN (c.f. \citealt{for14, gen14,coi14, aza17}). 

In order to quantify the expected contribution to the measured [\ion{O}{3}] emission line width from virial motions of the gas within the galactic potential, we calculate the ``expected velocity dispersion" following \citet{van15} using photometrically derived quantities. We use galfit \citep{galfit,galfit2} to model the $H$-band image using a Sersic profile + sky. The limited depth and ground-based seeing in the $H$-band image prevent a high-confidence fit; however, the best-fit models are disk-like, with semi-major $R_e=3.1$ kpc and axis ratio $q=0.3-0.4$ depending on whether we allow the Sersic index to vary or if we keep it fixed at $n=1$. We compute a circularized half-light ratio, 
\begin{equation}
\log(R_{e\rm{,c}})= \log R_e + 0.5 \log q
\end{equation}
yielding $R_{e\rm{,c}}=1.8-1.9$~{kpc}.  Next we calculate the expected velocity dispersion:
\begin{equation}
\log(\sigma_{\rm{predict}})=0.5(\log M_{\rm{star}} -\log R_{e\rm{,c}} -5.9  ),
 \end{equation}
where $\sigma_{\rm{predict}}$, $M_{\rm{star}}$, and $R_{e\rm{,c}}$ have units of \kms, \msun, and kpc, respectively. The predicted velocity dispersion of J1439B is $\sigma_{\rm{predict}}=190-200$~\kms, roughly 100 \kms\ less than the spectroscopically measured velocity dispersion. Given this, below we explore the likelihood that the kinematics of the gas reflect the dynamical properties of the stars within the galaxy or that of an AGN-driven wind.

\subsubsection{Broad Emission Lines in Compact Galaxies}

 A decade ago, deep HST imaging uncovered a significant population of massive compact galaxies at high redshift (c.f. \citealt{tru06, van08}). More recently, the remarkably concentrated nature of these sources was confirmed with stellar velocity dispersion measurements \citep{van09,tof12,van13,bel14a, hil16,bar16,bel17}, many of which are comparable to or larger than that of J1439B. 

\begin{figure}
\center
\includegraphics[width=0.5\textwidth]{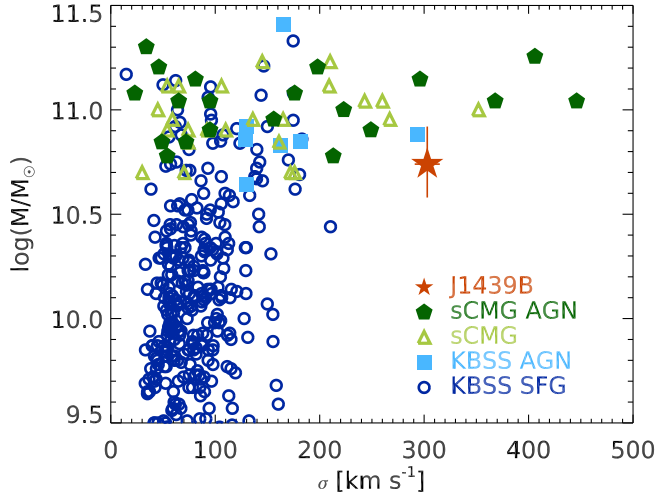}
\caption{Ionized gas velocity dispersion---stellar mass relation for $z\simeq2$ galaxies. [\ion{O}{3}] emission-line velocity dispersions from the UV-selected star-forming galaxies from the KBSS \citep{str17} are plotted in blue, with those that are identified as AGN in light blue squares. H$\alpha$-based velocity dispersions from the star-forming compact massive galaxies (sCMGs) described in \citet{van15} are shown in green, with X-ray-detected AGN shown as the dark green pentagons. The galaxy J1439B has a high measured $\sigma=303 \pm 12$~\kms, higher than all of the measured $\sigma$ of the KBSS. Even among the compact star-forming sample, the majority of galaxies with $\sigma>300$~\kms~have X-ray-detected AGN. }
\label{m-sigma}
\end{figure}

Van Dokkum \citeyearpar{van15} targeted a sample of plausible star-forming progenitors of these systems, characterized by compact sizes, large stellar masses, and high star-formation rates. The authors followed up galaxies that had photometrically predicted velocity dispersions, $\sigma_{\rm{predict}}>250$ \kms, finding spectroscopic velocity dispersions measured in the ionized gas that vary greatly from system to system.  Specifically, \citet{van15} obtained $K$-band spectroscopy for 25 star-forming compact massive galaxies (sCMG) and measured the velocity dispersion using H$\alpha$ emission (see Figure \ref{m-sigma}). Even among this sample selected on properties expected to obtain high velocity dispersions, only five galaxies have measured emission-line velocity dispersions $\sigma \simeq 300$~\kms (comparable to that of J1439B), four of which are X-ray-detected AGN with $L_{\rm{X}}>10^{43}$ erg s$^{-1}$. Van Dokkum \citeyearpar{van15} argued that the velocity dispersion of those systems are likely affected by the presence of the AGN, either through winds or the dynamics of gas close to the black hole\footnote{Note that broad H$\alpha$ emission could plausibly originate in the BLR close to a black hole; however, collisionally excited forbidden emission such as [\ion{O}{3}] is not expected from a BLR due to the higher gas densities (see, e.g., \citealt{sul00}).}.

Without a measurement of the stellar velocity dispersion in J1439B, one cannot rule out the possibility that the measured [\ion{O}{3}] emission line width is due to virialized motion alone. Nonetheless the argument above, combined with the high incidence of AGN among the studied population of star-forming galaxies with $\sigma \simeq 300$~\kms, further suggests that J1439B contains an AGN and that the kinematics of the ionized gas may be due to a wind.
 
\subsubsection{Broad Forbidden-line Emission due to an AGN-driven Outflow}

\label{outflow}

Broad emission lines in the optical spectrum of a galaxy are often an indication of the presence of an AGN. Typically, the broadest line emission is found in the Balmer lines and originates within the broad line region (BLR), thought to be a disk of gas in orbit around the black hole. Such regions do not exhibit collisionally excited emission lines such as [\ion{N}{2}] and [\ion{O}{3}] \citep{sul00}, presumably because the densities within the disk are too high to produce forbidden emission. Given the poor fit to the region surrounding H$\alpha$  and very low level emission at H$\beta$ in the J1439B spectrum (see Section \ref{spec_fit} and Figure \ref{fit}), very broad Balmer line emission cannot be ruled out; however, the relatively high velocity dispersion measured for the [\ion{O}{3}]$\lambda5008$ emission must also be explained and does not originate from a BLR.

\citet{for14} and \citet{gen14} reported the high incidence of broad emission components associated with the nuclei of $z\simeq2$ high-mass star-forming galaxies, which they interpret as signatures of AGN-driven outflows. The broad components are measured in recombination emission (H$\alpha$) as well as forbidden transitions ([\ion{N}{2}] and [\ion{S}{2}]), suggesting they are more likely due to winds than the BLR. \citet{gen14} report that two-thirds of galaxies with $\log(M/$\Msun$)\gtrsim10.9$ exhibit nuclear emission with line widths in the range of $190 \lesssim \sigma \lesssim 2250$ \kms. \citet{snew12} found that lower-mass star-forming galaxies also often exhibit underlying broad forbidden and recombination emission, which they also interpret as being due to outflows; however, this emission appears to extend across the whole galaxy, has lower typical line widths $\langle\sigma\rangle \simeq 160$ \kms, and is more likely driven by intense star formation. As J1439B has a relatively low SFR, if the broad [\ion{O}{3}] emission is due to winds, it is unlikely that the winds are driven by star formation alone. 

\citet{gen14} also found that the width of the broad nuclear emission appears to correlate with the position of the galaxy above or below the star forming main sequence. Specifically, \citet{gen14} stacked the nuclear spectrum of galaxies in three bins of stellar mass and two bins of SFR (above and below the main sequence). For their stack of galaxies with $10.6<\log(M/$\Msun$)<10.9$ with measured SFR that places them below the main sequence, consistent with the photometrically derived properties of J1439B, the average measured velocity dispersion of the broad nuclear component is $\sigma\simeq290$ \kms, very similar to the measured line width of the [\ion{O}{3}] emission in the FIRE spectrum of J1439B.

Given the high frequency of AGN in galaxies with masses comparable to J1439B and the likelihood that the large measured emission-line velocity dispersion of J1439B is caused by outflowing winds and the lower-than-typical SFR of J1439B, one must consider the possibility that DLA$_{\rm{J1439}}$ was expelled from the ISM of J1439B in a previous episode of feedback or through stripping. 

\begin{figure}
\center
\includegraphics[width=0.45\textwidth]{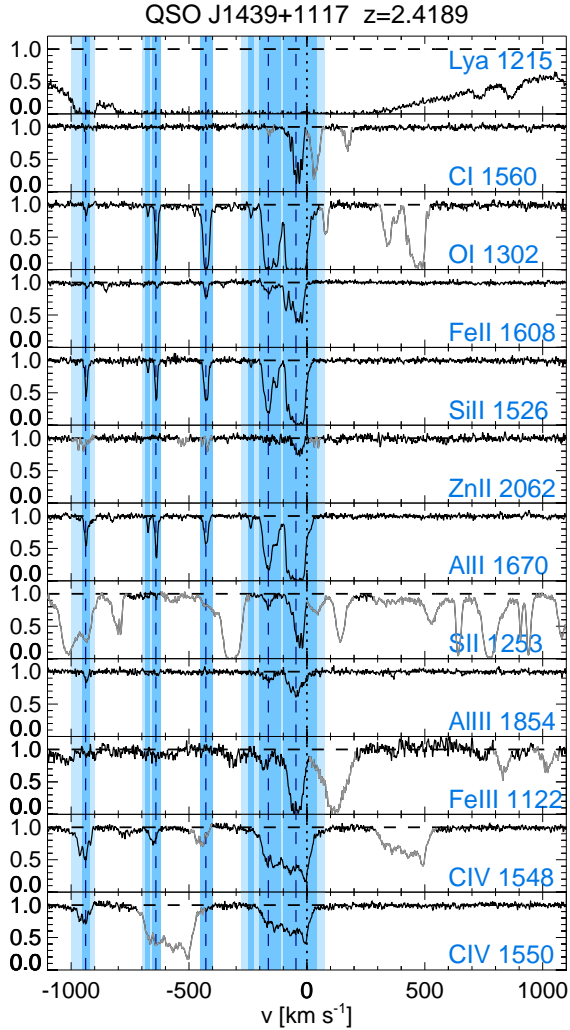}
\caption{Continuum-normalized QSO spectrum centered on the redshift of the galaxy J1439B. Portions of the spectrum that are contaminated by metal lines from gas at a different redshift are plotted in gray. In light blue shading are highlighted the spectral regions with high-ionization absorption (\ion{C}{4}); in the darker shade are highlighted the regions with low-ionization metal-line absorption. High-ionization absorption in \ion{C}{4} is strongest at the redshift of J1439B. The center of the sub-DLA and the highest column density low-ionization absorption is at $-46$ \kms~ with respect to J1439B. The locations of the five $\log(\NHI)>17.2$ absorbers used to fit the \ion{H}{1} distribution in \citet{sri08} and \citet{not08}, including the location of the sub-DLA, are marked with vertical dashed lines. High-ionization absorption is found at velocities $-1000<v<75$ \kms~and low-ionization absorption is found over the range $-950<v<40$ \kms.}
\label{stack}
\end{figure}

\begin{figure}
\center
\includegraphics[width=0.5\textwidth]{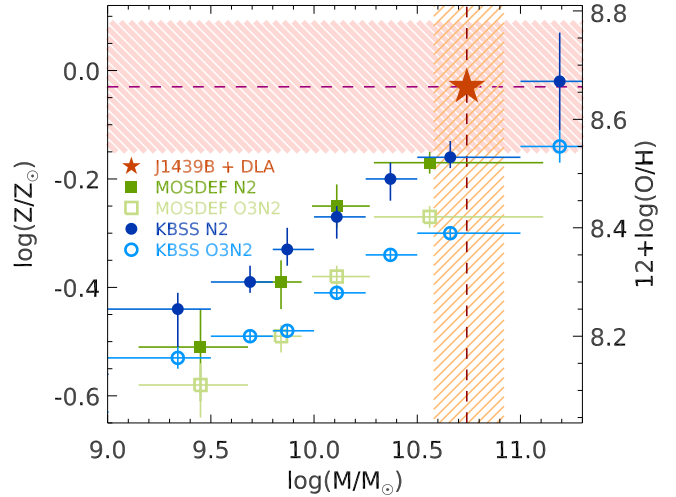}
\caption{Mass---metallicity relation (MZR) for $z\simeq2$ galaxies.  The range of [S/H] metallicity (a proxy for [O/H]) measured for DLA$_{\rm{J1439}}$ is shown as the horizontal line, and the pink shaded horizontal region is the error from \citet{not08}. For comparison, also shown are masses and $12+$log(O/H) (right axis) for two samples of star-forming galaxies with comprehensive NIR spectroscopy. In blue is shown the MZR for the KBSS sample as presented in \citet{ccs14} using the N2 (dark blue filled circles) and O3N2 (light blue open circles) strong line metallicity indicators. In green squares are shown the analogous data from the MOSDEF survey as presented in \citet{san15}. While there still exists considerable uncertainty in the MZR at $z\simeq2$, it is clear that solar metallicity gas is uncommon in systems with $\log(M/M_\odot)\ll 11$. The best-fit and range of stellar masses derived from the photometry of galaxy J1439B is shown in the orange shading and vertical dashed line. Combining the metallicity of DLA$_{\rm{J1439}}$ and the stellar mass of J1439B (assuming the sub-DLA represents the ISM of the galaxy) would give the red star. The location of the data point for J1439 in comparison to the MZR of the $z\simeq2$ samples is consistent with the possibility that the sub-DLA gas could be ejected ISM from J1439B. }
\label{MZR}
\end{figure}

\subsection{The Metallicity of DLA$_{\rm{J1439}}$ and Implications for Its Relationship to Galaxy J1439B}

\label{MZR_text}

There are many remarkable properties of DLA$_{\rm{J1439}}$, but one of the most discriminatory as to the likely origin of the gas is its metallicity. Metal line absorption associated with the sub-DLA is detected from a wide variety of elements and arises from gas with a large range of ionization states (see Figure \ref{stack}). As the \ion{O}{1} absorption associated with the trough of the sub-DLA is saturated, a direct determination of the O/H ratio is not possible. However, \citet{not08} reported a sulfur abundance based on fits to the \ion{S}{2} absorption components of [S/H]$=-0.03\pm0.12$. The ratio of S/O is constant across galactic environments \citep{pag09}, and so sulfur can be used as a proxy for oxygen abundance. 

There is considerable uncertainty regarding the true oxygen abundance of the ISM of high-redshift galaxies. Because oxygen abundances can only be directly measured for a rare subset of galaxies at high-$z$ \citep{yua09, wuy12, chr12,bay14,san16b}, most of which are gravitationally lensed, the majority of the inferences regarding the chemical abundances of distant galaxies are based on the ratios of ionization and excitation-sensitive emission-line ratios such as N2, O3N2, and R$_{23}$ (see, e.g., \citealt{pet04}). Unfortunately, these metallicity diagnostics introduce systematic uncertainties as large as 0.7 dex \textit{even in the local universe} \citep{kew08}. Further, the typical emission line ratios of star-forming galaxies evolve with redshift  (see Section \ref{BPT_text}), which manifest as large discrepancies in the abundances derived \textit{for the same high-$z$ galaxy} using different strong line diagnostics \citep{ccs14,san15}. Thankfully, even given these considerable caveats, the high level of enrichment of the sub-DLA is constraining because all determinations of the high-$z$ mass-metallicity relation suggest that solar metallicity enrichment only occurs in very massive galaxies. 

Figure \ref{MZR} shows the most recent determinations of the mass-metallicity relation of $z\simeq2$ star-forming galaxies based on statistical samples with uniform data. We consider metallicities determined from both the N2 (filled symbols) and O3N2  (open symbols) indicators as measured in the $H$-band-selected MOSDEF sample (\citealt{san15}, green squares) and the UV-selected KBSS sample (\citealt{ccs14}; blue circles). For comparison, in pink shading we show the constraints on the $\alpha$ abundance of the sub-DLA. Clearly, whichever galaxy responsible for the absorption system must have a high stellar mass capable of producing roughly solar metallicity enrichment. The mass of J1439B, indicated by the orange vertical shading, is consistent with the possibility that the sub-DLA is composed of gas that originated within the ISM of J1439B. Recent work studying the relationship of the masses of DLA galaxy counterparts and their measured metallicities also suggests that the mass of the galaxy associated with DLA$_{\rm{J1439}}$ would be $\log(M/M_\odot) >10.5$, consistent with the mass of J1439B \citep{mol13, chr14}.

\section{Discussion}

\label{discussion}

While many DLAs and sub-DLAs likely originate within the central gaseous regions or extended disks of galaxies and may therefore commonly trace the \textit{in situ} ISM of the galaxy, the large separation between DLA$_{\rm{J1439}}$ and J1439B (38 pkpc) makes this scenario improbable. Even in the unlikely possibility that such a large gas disk or other coherent structure existed surrounding J1439B, extending 10-20 times farther than the optical light of the galaxy, we would expect the chemical enrichment of the gas at such a large distance to be significantly lower than that within the central portions of the galaxy (c.f. \citealt{ken03,zur12,bre16}). Given that a disk origin of DLA$_{\rm{J1439}}$ is disfavored, we must consider the possibility that the sub-DLA originates from either another galaxy closer to the line of sight or from ejected or stripped ISM from J1439B.

\subsection{Could Another Galaxy be Responsible for DLA$_{\rm{J1439}}$?}

\label{another_gal}

As outlined above, there is considerable evidence of a plausible connection between the galaxy J1439B and the DLA$_{\rm{J1439}}$. However, if J1439B is the galaxy counterpart to the sub-DLA, it is at a larger separation than is typically found between high-\NHI\ absorbers and their host galaxies (\citealt{fum15}, but see also \citealt{nee17}). Here we explore the possibility that another galaxy could be responsible for the sub-DLA. 

In Section \ref{MZR_text} and Figure \ref{MZR}, we showed that the sub-DLA's high metallicity implies that the source of its enrichment is likely a massive galaxy with mass comparable to or greater than that of J1439B. The images shown in Figure \ref{images} rule out the existence of another galaxy with comparable mass  to J1439B lying closer to the line of sight to the QSO, unless the galaxy is hidden within the PSF of the QSO.\footnote{The $H$-band image of J1439B provides a 14$\sigma$ detection of the galaxy. As this corresponds to rest-frame $\sim5000$\AA, the luminosity is a good proxy for mass. We would detect a galaxy of comparable size but five times less luminous at 3$\sigma$. The other source directly east of J1439B that is visually apparent in Figure \ref{images} has a photometric redshift computed with EAZY \citep{eazy} that is inconsistent with solutions $z>2$. No other faint galaxies are detected with separations $<5"$ from the QSO.}

Additional constraints on the existence of a galaxy very close to the line of sight to the QSO were discussed in \citet{bou12} using data from the z2SIMPLE survey taken with the Spectrograph for INtegral Field Observation in the Near-Infrared (SINFONI; \citealt{sinfoni}) on the VLT. \citet{bou12} took $K$-band IFU spectroscopy in natural seeing of a 10'' $\times$ 10'' field surrounding 28 QSOs, including QSO J1439+1117. Among their 28 absorbers, they detected only 4 possible host galaxies and did not detect a galaxy at the redshift of DLA$_{\rm{J1439}}$.\footnote{The position of J1439B is very near the edge of the field of view of the SINFONI observations and only received half the exposure time.}

In the central 4" search area surrounding the QSOs, \citet{bou12} reported a $5\sigma$ limiting H$\alpha$ flux of $2.3\times10^{-17}$ \ergscm2, corresponding to a  $5\sigma$ limiting SFR of 3.9 \Msun~yr$^{-1}$ uncorrected for dust, integrated over eight spectral pixels, or roughly 300 \kms. For a galaxy analogous to J1439B with an emission line velocity dispersion $\sigma=300$~\kms, the true limit is somewhat less stringent as the line emission would be spread over a larger number of pixels. Nevertheless, the SINFONI data further demonstrate the lack of a galaxy closer to the line of sight. \citet{bou12} argued that only within a 0\farcs3 radius of the QSO would the SINFONI data been insensitive to the H$\alpha$ emission from the galaxy. 

The final possibility is that another galaxy lies directly on top of the line of sight to the QSO and has an unfavorable contrast ratio, leaving it undetected in the imaging and IFU observations. This is a challenging possibility to disprove, but we can quantify the likelihood in part using measurements of the pair fraction of massive $z\simeq2$ galaxies.  

There have been several studies aimed at quantifying the merger rate at high redshift that have measured the fraction of massive galaxies that are found with close companions.  In this work, we are most interested in pairs of nearly equal mass since, as we argue in Section \ref{MZR_text}, the metallicity of DLA$_{\rm{J1439}}$ indicates that the source of its metals is likely a galaxy of mass equal to or greater than that of J1439B. Given this, we consider the results of studies qualifying the pair fraction of 1/4 mass-ratio pairs to be an upper limit on the likelihood that another galaxy would be the primary origin of the absorber. Three recent studies have considered the incidence of satellites surrounding massive $\log(M/\msun) > 10.54 - 11$ galaxies with $1.5<z<3.0$ and projected separations less than $\sim40$ kpc, finding that $7\%-15\%$ of such galaxies have a likely companion at least one-quarter as massive as the primary \citep{new12,man12,man16}.  

In practice, we are only interested in the volume surrounding the galaxy that is closer to the line of sight to the QSO, and we have ruled out the existence of such a galaxy in the majority of that volume, but not at the precise location of the QSO line of sight. Further, we have argued that galaxies with  $\log(M/\msun) \ll 11$ are unlikely to be responsible for the high levels of chemical enrichment found in the sub-DLA. Given these caveats, we consider $7\%-15\%$ to be a very conservative upper limit on the probability that another galaxy is responsible for DLA$_{\rm{J1439}}$.

\subsection{Implications}
\label{QSO}

If DLA$_{\rm{J1439}}$ represents the ejected or stripped ISM of J1439B, it offers a rare window into the chemistry and molecular properties of the ISM in high-$z$ galaxies.  As shown in Figure \ref{stack}, the environment surrounding DLA$_{\rm{J1439}}$ and J1439B contains several strong multiphase absorbers. Collectively, these systems probe gas within the circumgalactic medium of J1439B. \citet{sri08} and \citet{not08} modeled the \ion{H}{1} absorption within 1000 \kms\ of the sub-DLA using four additional absorbing structures, all with $\log(\NHI)>17.2$ (noted in Figure \ref{stack} by the vertical dashed lines).  Absorbers with $\log(\NHI)>17.2$, commonly referred to as Lyman Limit Systems, are intrinsically rare, with a typical incidence of $\sim1$ per QSO sightline at high redshift \citep{sto94}. Therefore, the existence of five absorbers located within 1000 \kms\ of each other all with $\log(\NHI)\gg17.2$ is a rare occurrence and suggests a direct relationship with the sub-DLA and likely J1439B. 

We argue above that the metallicity of the sub-DLA itself is consistent with the enrichment expected for the ISM of J1439B. Given the existence of these other strong \ion{H}{1} systems, one can further constrain the likely origin of the gas. As mentioned above, the \ion{H}{1} absorbers are spread over $\sim900$ \kms. Each of these absorbers is also detected in metallic transitions with ionization potentials ranging from  \ion{O}{1} to \ion{C}{4}. The velocity distribution of metal line absorption is spread over $\Delta v > 1000$ \kms\ both in low- and high-ionization gas. Given the large velocity spread, we favor a wind scenario for their origin as the expected velocities associated with the gravitational potential of J1439B are significantly less than 1000 \kms. If we assume the stellar mass -- halo mass relation from \citet{beh13}, J1439B has a dark matter halo mass of $M_{\rm halo}\simeq10^{12.4} M_\odot$. Thus the virial velocity of the halo would be $\sim300$ \kms, far less than the velocity spread of absorbers seen in the QSO spectrum. This precludes an origin for the absorbers from either virialized gas within the CGM of the galaxy or stripped ISM resulting from an interaction. Therefore, we argue that DLA$_{\rm{J1439}}$ and its surrounding absorbers provide a unique view of AGN-driven winds at high redshift.

\subsubsection{The Chemical Evolution of J1439B}

The detailed elemental abundances of DLA$_{\rm{J1439}}$ provide insight into the nature of the ISM within the source galaxy.  The abundances of S and Zn, which are typically not depleted onto dust grains, are consistent with a high level of enrichment, roughly solar metallicity ($[\rm{S}/\rm{H}]=-0.03\pm0.12$, $ [\rm{Zn}/\rm{H}]=+0.16\pm0.11$), and their ratios can be used to measure the level of $\alpha$-enhancement. In DLA$_{\rm{J1439}}$ the gas is consistent with solar $\alpha$/Fe \citep{not08}. This is notable given the significant $\alpha$-enhancements that are common in the old stars within giant elliptical galaxies \citep{wor92,hen99,tra00,tho05}, and the recent finding of super-solar O/Fe [$\simeq 4 - 5$ (O/Fe)$_{\odot}$] in UV-selected star-forming galaxies at the same cosmic epoch \citep{ccs16}. Interestingly, \citet{not08} argue that the N/O ratio in DLA$_{\rm{J1439}}$ also appears to be consistent with the solar ratio, again somewhat at odds with the recent determination of [N/O]$ = -0.38\pm0.04$ in $L^*$ star-forming galaxies at $z\sim2$ from \citet{ccs16}. Both the $\alpha$/Fe and N/O ratios suggest a level of chemical maturity associated with prolonged star formation histories, consistent with the $\sim$Gyr age estimate of J1439B from the SED fitting. 

Comparison of refractory and volatile elemental abundances provide some constraint on the amount and type of dust in DLA$_{\rm{J1439}}$ \citep{pet97b,wol03}.  \citet{sha16} considered detailed models of DLA$_{\rm{J1439}}$ and noted that the dust-to-gas ratio inferred for the sub-DLA is dependent on which volatile element is used in the calculation. Employing the measured [Fe/Zn] abundance, they infer a dust-to-gas ratio nearly twice the value in the Milky Way, whereas when using [Fe/S] they find a ratio very similar to the Milky Way value. Regardless of which value is used, however, DLA$_{\rm{J1439}}$ is again atypical of damped absorbers which have a typical dust-to-gas ratio roughly 30 times lower than the Milky Way \citep{pet97b}. The large amount of dust in the system is consistent with its significant degree of chemical enrichment, and further supports the scenario in which the sub-DLA is related to a massive evolved galaxy such as J1439B. 


\subsubsection{AGN-driven Outflows}

\label{outflows2}

The origin of molecular gas in outflows is debated. Some theoretical arguments favor the direct expulsion of molecular gas through energy-conserving winds \citep{fau12,zub14,tom15}, while in other models the molecular gas forms \textit{in situ} in dusty outflows due to cooling instabilities \citep{fer16}. Regardless of their origin, however,  outflows containing molecular gas have been definitively observed in several local AGN and QSOs \citep{fer10,ala11,stu11}, extending in some cases over many kiloparsecs \citep{cic14}. Extended outflows have also been observed in multiple phases, both in ionized \citep{gre12,liu13a,liu13b,liu14,hai14,sun17,yum17} and [\ion{C}{2}]-emitting gas \citep{cic15,mai12} at both high and low redshift.  \citet{har14} considered a sample of 16 $z<0.2$ type 2 AGN, finding that all of them had ionized gas emission lines similar to those observed in J1439B with $\sigma>300$ \kms, and that 70\% of such galaxies have outflows that extend over kpc scales. 

Interestingly, many authors have noted a relationship between the detected extent of the ionized outflow and the luminosity of the AGN \citep{liu13a,liu13b,liu14,hai14,sun17}. In particular, extended ionized outflows seem to be most common in AGN with $L_{\rm{bol}} \gtrsim 10^{46}$ erg s$^{-1}$ \citep{sun17}. Following, \citet{rey08} and \citet{liu09}, who derived a relationship between the bolometric luminosity of AGN and the luminosity of their [\ion{O}{3}] emission, we calculate $L_{\rm{J1439B,~[O~III]}}=10^{42.6} $ erg s$^{-1} = 10^{9.0} L_\odot$, which would imply $L_{\rm{J1439B,~bol}}=10^{46.0}$ erg s$^{-1}$. This suggests that the AGN in J1439B is comparably luminous to the class of lower-redshift AGN that have been shown to drive outflows. 

While clearly uncertain, it is plausible that the DLA$_{\rm{J1439}}$ represents wind material from an AGN-driven outflow from J1439B. In that case, the J1439 DLA-galaxy system offers many unique clues about the nature of such a wind. In particular, given the detection at 38 kpc, J1439 provides spatial information about the extent of the wind through a more sensitive probe than in emission. Additionally, the properties of the gas can be studied in much greater detail through the sub-DLA's absorption lines, providing high-fidelity measurements of the chemistry, density, and kinematics of the gas at large distances from the galaxy.  

Given the measurement of $\log(N_{\rm{CO}})=13.89\pm0.02$  and $\log(N_{\rm{H}_{2}})=19.38\pm0.10$ in this sub-DLA by \citet{sri08}, it is possible to measure $N(\rm{CO})/N(\rm{H}_2)=3\times10^{-6}$.  This is significantly less than the value typical of star-forming regions in the local universe, where $N(\rm{CO})/N(\rm{H}_2)=2.7 \times 10^{-4}$ \citep{lac94}.  The ratio is comparable to values observed in diffuse clouds in the Galaxy, but at fixed $\log(N_{\rm{H}_{2}})$, DLA$_{\rm{J1439}}$ has 0.5-1 dex more CO \citep{bur07, she08}. Using the ratio of fine-structure carbon absorption lines, \citet{sri08} also derived a density of the sub-DLA to be $45-62$ cm$^{-3}$, comparable to, although somewhat lower than, the typical densities in diffuse clouds within the ISM of the Milky Way \citep{sno06}. Taken together, these two measurements suggest that the physical properties of the gas are likely similar to that of diffuse Galactic clouds. 

In addition to the properties of the gas and their implications for its physical state, the kinematics of the various absorption components are also of interest. In Figure \ref{stack}, one can clearly see a number of distinct metallic absorption features. While the redshift of the galaxy $z=2.4189$ is well aligned with the strongest components of the neutral and low-ionization gas ($z=2.41837$, $\Delta v=46$ \kms), the majority of the metal line absorption systems lie blue-shifted with respect to both the sub-DLA and J1439B, with the highest-velocity absorbers detected in low-ionization (e.g. \ion{O}{1}, \ion{Si}{2}) as well as high-ionization (e.g. \ion{C}{4}) lines extending to $\sim950$ and $\sim1000$ \kms\ respectively. Redshifted metal lines extend to less than 100 \kms\ in the opposite direction. 

As we argue above, the large velocity spread of the metal-line absorbers detected in the QSO spectrum are inconsistent with a gravitational origin, given the mass of J1439B. The most plausible scenario is that the absorbers result from a wind with an outflow velocity $v \gtrsim 1000$ \kms, again suggestive of AGN driving. In the context of an outflow model with a constant outflow velocity, the kinematics of the absorbing gas provide a joint constraint on the opening angle of the outflow and the outflow velocity. For an outflow velocity of 3000 \kms~or less, the opening angle would be larger than 20 degrees. 

The presence and high fraction of molecular gas within the sub-DLA [$f=2N(\rm{H}_2)/(N(\rm{H}~ I)+2N(\rm{H}_2))=0.27^{+0.10}_{-0.08}$; \citep{sri08}] is also of considerable interest. In particular, if the absorbing gas was ejected from J1439B in its present state (as opposed to formed within the outflow), it would provide a possible explanation for the galaxy's anemic SFR, given that the direct fuel for star formation would have been removed. 

In addition to this, the most likely scenario is that this gas escapes completely from the halo. As mentioned above, we assume the stellar mass -- halo mass relation from \citet{beh13} to infer that J1439B has a dark matter halo mass of $M_{\rm halo}\simeq10^{12.4} M_\odot$. The measured velocity range in the metal line absorbers ($v \simeq 1000$ \kms) is comparable to the escape velocity of the halo at a distance of 38 kpc.  So if DLA$_{\rm{J1439}}$ is the ejected ISM of J1439B, it represents the permanent removal of star-formation-ready material from a galaxy, a very effective and long-lasting form of feedback. 

\subsubsection{Timescales}

Lastly, we comment briefly on the timescales of interest implied by the inferred velocities of the wind and the distance between J1439B and the QSO line of sight. In Section \ref{outflow}, we argue that the large measured emission-line velocity dispersion of J1439B is most likely due to an AGN-driven wind. This suggests a lower limit for the current outflowing wind speed of $\sim$ 300 \kms. Above in Section \ref{outflows2}, we  discuss the possibility of a $\gtrsim$1000 \kms\ outflow based on the kinematics of the gas detected in absorption. Noting the projected distance between the galaxy and the QSO line of sight of 38 kpc, it would take $\sim10^8$ years for gas traveling at 300 \kms\ to cover this distance. If that gas were instead ejected at 1000 \kms, it would require only $4\times 10^7$ years. Either of these timescales is significantly shorter than the typical age of the stars suggested by the SED fit ($\sim$700 Myr). This is consistent with the high levels of enrichment measured in the sub-DLA; a majority of the stars in the galaxy would have formed prior to the ejection of the gas, enriching the ISM to the solar metallicity value measured. Further, the travel time required is similar to the expected lifetime of AGN in the distant universe \citep{hai98,hae98b,yu02,mar04,hop09}.

\section{Summary}

We report the discovery of a galaxy 38 kpc from the sightline to QSO J1439+1117 coincident with the redshift of the CO-bearing, solar-metallicity sub-DLA. We presented optical and NIR photometry of the galaxies as well as SED fits showing that the galaxy is massive ($\log(M/M_\odot)=10.74^{+0.18}_{-0.16}$) and evolved (Section \ref{SED_text} and Figure \ref{sed}). A NIR spectrum of the galaxy, J1439B, observed with \textit{Magellan}/FIRE, shows broad [\ion{O}{3}] emission ($\sigma=303\pm12$ \kms) as well as [\ion{N}{2}] and H$\alpha$ emission (Section \ref{spec_fit} and Figure \ref{fit}). 

Based on the properties of the galaxy and the sub-DLA, we argue that there is a plausible connection between J1439B and DLA$_{\rm{J1439}}$:
\begin{itemize}
\item{The galaxy has a lower-than-typical SFR given its stellar mass, suggesting that there could be some process occurring, or that occurred in the past, that prevents star formation (Section \ref{MSSF_text} and Figure \ref{MSSF}).}
\item{The ionization properties of J1439B determined from the  [\ion{N}{2}] BPT and MEx diagrams favor the galaxy harboring an AGN (Section \ref{BPT_text} and Figure \ref{BPT}).} 
\item{The large velocity dispersion observed in J1439B's nebular emission lines is most plausibly explained by an AGN-driven outflow (Section \ref{width_text} and Figure \ref{m-sigma}).}
\item{The metallicity of DLA$_{\rm{J1439}}$ requires enrichment within a high-mass, evolved galaxy, similar to J1439B (Section \ref{MZR_text} and Figure \ref{MZR}).}
\end{itemize}

In Section \ref{another_gal}, we explored the possibility that another galaxy might lie closer to the line of sight and be the true galactic counterpart to the sub-DLA. Based on our imaging data, as well as published IFU spectroscopy, we conclude that another massive galaxy does not lie closer to the line of sight, unless it is coincident with the QSO sightline. Based on pair statistics for massive high-redshift galaxies, we conclude that there is no more than a 15\% chance that another massive galaxy lies within $\sim$40 kpc of the line of sight. 

We conclude that J1439B is likely the galactic counterpart of DLA$_{\rm{J1439}}$, and that the sub-DLA is most likely the ISM ejected by the AGN. In this scenario, the properties of DLA$_{\rm{J1439}}$ offer a rare view of the chemistry of the ISM close to a high-redshift AGN, and of the state of gas driven out by AGN feedback. The sub-DLA has solar $\alpha$/Fe and N/O abundance ratios, suggesting that the galaxy is quite chemically evolved with a longer star formation history. The molecular properties of the sub-DLA are most similar to that of diffuse clouds in the Milky Way, and the dust-to-gas ratio is similar to or larger than that of the Milky Way.  Metal line absorption is detected with a large velocity spread ($>1000$ \kms), implying large outflow velocities if the gas results from a wind from J1439B. At a distance of 38 kpc, the large velocity spread also suggests much of the absorbing material would be unbound from J1439B. Given the significant molecular gas content within the sub-DLA, this would represent the ejection of gas that otherwise would have likely undergone star formation. 

The collective properties of J1439B and DLA$_{\rm{J1439}}$ are consistent with the picture in which AGN play an active role in modulating the star formation of massive, high-redshift galaxies.  Confirmation of the X-ray and radio properties of J1439B would provide better understanding of the nature of the AGN, and the detection of extended molecular emission would cement the connection between J1439B and DLA$_{\rm{J1439}}$. These further observations provide a unique opportunity to obtain a more detailed view of AGN feedback in the high-redshift universe.

\acknowledgements

The authors wish to thank D. Kelson for the use of his FourCLift FourStar Reduction code and for his assistance with it. We acknowledge insightful discussions with J. Grenne, D. Kelson, J. Simon, S. Tonnesen, and C. Steidel which improved the content of the paper. We also thank J. Kollmeier and S. Dong for obtaining the IMACS optical photometry of the galaxy, I. Shivaei and N. Reddy for providing the MOSDEF data plotted in Figure \ref{MSSF}, and A. Strom for providing catalogs of the KBSS measurements. The authors thank the anonymous referee whose comments improved the manuscript.

M.T.M. thanks the Australian Research Council for Discovery Project grant DP130100568 which supported this work. This paper includes data gathered with the 6.5 m Magellan Telescopes located at Las Campanas Observatory, Chile as well as observations collected at the European Organisation for Astronomical Research in the Southern Hemisphere under ESO program 278.A-5062(A) and obtained from the ESO Science Archive Facility. 

\scriptsize
\bibliographystyle{apj}
\bibliography{J1439}

\end{document}